\documentclass[twocolumn,prc,showpacs]{revtex4}
\usepackage{graphicx}

\begin{document}

\bibliographystyle{apsrev}

\title{Thermal shape fluctuation effects in the description of hot nuclei}
\author{V. Martin}
\affiliation{An\'alisis Num\'erico, Facultad de Inform\'atica, Universidad Polit\'ecnica de Madrid,
28660 Madrid, Spain}

\author{J.L. Egido}
\author{L.M. Robledo}
\affiliation{Departamento de F\'{\i}sica Te\'orica C-XI, Universidad Aut\'onoma
de Madrid, 28049 Madrid, Spain}
\date{\today}

\begin{abstract}
The behavior of several nuclear properties with temperature is analyzed within the framework
of the Finite Temperature Hartree-Fock-Bogoliubov (FTHFB) theory with the Gogny force and 
large configuration spaces.  Thermal shape fluctuations in the quadrupole
degree of freedom, around the mean  field solution, are taken into account with the Landau 
prescription. As representative examples the nuclei $^{164}$Er,  $^{152}$Dy 
and $^{192}$Hg are studied.  Numerical results  for the superfluid to normal and deformed to 
spherical shape  transitions are presented.  We found a substantial effect of the
 fluctuations on the average value of several observables. 
 In particular, we get a decrease in the critical temperature ($T_c$) for the shape transition 
as compared with the plain FTHFB prediction as well as a washing out of 
the  shape transition signatures.
The new values of $T_c$ are closer to the ones found in Strutinsky calculations and with the 
Pairing Plus Quadrupole model Hamiltonian.

\end{abstract}

\maketitle

\section*{Introduction}

Since the advent of the new generation of \( 4\pi  \) gamma ray detectors and the
improved accuracy in the channel selection new possibilities have opened up
in the study of nuclear structure. 
Besides this, the availability of faster computers has
made possible to perform realistic theoretical investigations with large configuration spaces. 
The high excitation energy  is specially interesting since new features may take place. 
For example, in the quasicontinuum,  the high level density gives rise to the unexpected 
phenomenon of the damping of the rotational motion. In the limit of high excitation energies
(or temperature $T$) quantum effects become less relevant or may even disappear. 
 Thus one expects that in a heated nucleus physical effects like 
superfluidity or shape deformations are washed out when $T$ increases.
This expectation can be easily understood in terms of the shell model since, by increasing 
$T$, one promotes particles  from levels below the Fermi 
surface to levels above it. In the case of pairing correlations, blocking levels
amounts to destroying Cooper pairs. In the case of shape deformation, by 
depopulating the deformation driving levels (intruders)  one gets on the average less deformation.
Experimental information about nuclear shape changes can be obtained  
by  means of the Giant Dipole Resonance (GDR) built on excited states.
Exclusive experiments studying the GDR strength at a given  excitation 
energy (or $T$) of the nucleus have been carried out in 
refs.~\cite{AGH.90,NBH.92,NBH.99,HBB.03}. The understanding of these phenomena is relevant
because it affects important features like the fission barriers  and
the stability of the nucleus itself. For a recent review on hot nuclei see ref. \cite{ER.93}.

The shape transitions have been object of many studies, most of them with {\em schematic models},
separable forces, and {\em small configuration spaces} \cite{Mo.73,Goo.81,ERM.86,Goo.86,RR.94}.
The theoretical approaches used in the calculations are based on the mean
field approximation, mainly the Finite Temperature Hartree-Fock-Bogoliubov 
theory (FTHFB). The mean field approximations predict sharp shape transitions, whereas
for finite systems, however, one expects washed out transitions instead.
The fact that the predicted critical temperatures are rather high (around 2-3 MeV) indicates
that not only the most probable deformation is relevant but that there is a finite (in some
cases very large) probability for the system to have other shapes which should be taken into account. 
Calculations beyond mean field including  thermal fluctuations  have confirmed the expectation 
of washed out transitions \cite{EDR.86,8}.
 
  Theoretical studies with {\em effective forces} 
and {\em large configuration spaces} have been performed at the FTHFB level with 
density dependent forces, Skyrme \cite{BQ.74a,BQ.74b,QF.78} and recently with the Gogny
 \cite{EMR.00} force.  Additional calculations have been done in the relativistic mean field (RMF) 
approximation \cite{AT.00,GML.00}. Calculations including thermal fluctuations in conjunction 
with large configurations spaces and effective forces have been performed only
very recently \cite{AT.00,MER.03}.

  From these studies a discrepancy  has emerged since, while the {\em mean field approaches} (FTHFB) 
with {\em effective forces} (like Skyrme, the Gogny force or the relativistic approaches), provide
the view of a sharp shape transition at a relatively high critical temperature 
($T_c \approx 2.7$ MeV for $^{164}$Er),  schematic models (like the Pairing plus 
Quadrupole) and Strutinsky calculations provide also a sharp transition though at a much lower 
critical temperature  ($T_c \approx 1.7$ MeV  for $^{164}$Er). Furthermore, the discordant point 
of a "sharp" transition for a small system, like the nucleus, predicted by both approaches requires
further investigation.  Earlier calculations with the Pairing plus Quadrupole Hamiltonian 
\cite{ERI.85,EDR.86} have  pointed out the relevance of including fluctuations in mean field 
approaches at finite temperature as a step forward to clarify some aspects of these problems. It is the aim of 
this paper to investigate the problems just mentioned as well as other related high excitation 
energy topics, level densities, etc., within a beyond mean field theory.  Towards this end, 
the FTHFB calculations of ref.~\cite{EMR.00} with the Gogny force and large configuration
spaces will be generalized to include fluctuations in the quadrupole moment degree of freedom.

 The finite range density dependent  Gogny force has the advantage of providing the particle-hole
(Hartree-Fock) and the particle-particle (pairing)  matrix elements from the same interaction,
at variance with relativistic theories and most Skyrme calculations.
 In the  fitting of the D1S~\cite{1} 
parametrization  no excited states or spin dependent data was used, however, it has produced
good results in the description of nuclear properties not only at zero~\cite{2,3}
but also at large spin~\cite{254NoSpin} and, more recently~\cite{EMR.00}, in calculations
at high excitation energy.
Since our purpose is to study the behavior of shell effects and fluctuations with
temperature, we have selected both theoretically and experimentally well 
known nuclei that display a variety of shapes in the ground state:  strongly
deformed (\( ^{164} \)Er), oblate (\( ^{192} \)Hg ) and rather soft   (\( ^{152} \)Dy).

\section*{Theory}

To study the behavior of nuclei with increasing temperature we use the D1S~\cite{1}
parametrization of the finite range density dependent Gogny force~\cite{2,3} in the FTHFB 
framework~\cite{12,Goo.81}. 
The Gogny force, at variance with most of the Skyrme parametrizations and the relativistic models, 
allows full selfconsistent calculations since it provides the particle-hole
and pairing fields from the same force. 

 At finite temperature, as at temperature zero, the basic approximation is the
mean field theory. Its most sophisticated version, the FTHFB, has been
developed in refs.~\cite{6a,Goo.81,ERM.86}. For convenience we will give here a short
outline.

For a system  at constant temperature $T$ and with chemical potential $\mu$, the
equilibrium state can be obtained from the variational principle over the grand
canonical potential
\begin{equation}
\Omega =E-TS-\mu N.
\label{gcp}
\end{equation}
The energy, $E$, entropy, $S$, and particle number, $N$, are thermal averages defined
by
\begin{equation}
\begin{array}{lllll}
E & \equiv & \left\langle \hat{H}\right\rangle_T  & = & Tr(\hat{D}\hat{H}),\\
S & \equiv & \left\langle -k\ln \hat{D}\right\rangle_T  & = & -k \,Tr(\hat{D}\ln \hat{D}),\\
N & \equiv & \left\langle \hat{N}\right\rangle_T  & = & Tr(\hat{D}\hat{N}),
\end{array}
\end{equation}
with \( Z \) the grand partition function  and \( \hat{D} \) the density operator 
given by
\begin{equation}
\begin{array}{ccc}
Z & = & Tr[\exp (-\beta (\hat{H}-\mu \hat{N}))],\\
\hat{D} & = & Z^{-1}\exp (-\beta (\hat{H}-\mu \hat{N})),
\end{array}
\end{equation}
with \( \beta =1/kT \). 

  In the FTHFB approach  the density operator is approximated by 
\begin{equation}
   \hat D_0~=~{ {e^{\hat{\cal H}/T}} \over Z_0},
\label{E2.8}
\end{equation}
where $\hat{\cal H}$ is the most general  Hermitian single particle
 operator, to be determined by the variational principle and $Z_0$ is
the partition function. It can be shown \cite{ER.93} that ${\cal H}$ is given
by
\begin{equation}
   {\cal H}~=~\left(\begin{array}{cc}h&\Delta\\-
\Delta^*&-h^*\end{array}\right),
\end{equation}
with \( \Delta  \) the pair potential  and \( h \) the HF hamiltonian. \( h \)
is given in terms of the kinetic energy \( t  \), the
HF field, \( \Gamma  \), and the chemical potential,  \( \mu  \),
\begin{equation}
\begin{array}{ccc}
h & = & t +\Gamma -\mu, \\
\Gamma _{ij} & = & \sum _{kl}v_{ikjl}\rho _{lk},\\
\Delta _{ij} & = & \frac{1}{2}\sum _{kl}v_{ijkl}\kappa _{kl}.
\end{array}
\end{equation}
The density matrix, \( \rho  \), and the pairing tensor, \( \kappa  \),
are given by
\begin{equation}
\begin{array}{ccc}
\rho  & = & UfU^{+}+V^{*}(1-f)V^{t},\\
\kappa  & = & UfV^{+}+V^{*}(1-f)U^{t},
\end{array}
\end{equation}
and
\begin{equation}
f_{i}=\frac{1}{1+e^{\beta E_{i}}}.
\end{equation}
The matrices $(U,V)$ provide the relation between the quasiparticle 
and the single particle basis~:
 \begin{equation}
   \alpha^+_m~=~\sum_k U^{}_{km} c^+_k + V_{km} c^{}_k.
\end{equation}
 They are determined, together with the  quasiparticle energies, \( E_{i} \),
by the FTHFB equation
\begin{equation}
\label{hfbeq}
\begin{array}{lllll}
\left( \begin{array}{cc}
h & \Delta \\
-\Delta ^{*} & -h^{*}
\end{array}\right)  & \left( \begin{array}{c}
U_{k}\\
V_{k}
\end{array}\right)  & = & \left( \begin{array}{c}
U_{k}\\
V_{k}
\end{array}\right)  & E_{k}.
\end{array}
\end{equation}

The solution of this equations provides us with the configuration that minimizes
the grand canonical potential. With $U, V$ and $f$ known one can determine the density
operator $\hat{D}_0$ and calculate any expectation value.  For density dependent forces
like Skyrme or Gogny, the formalism remains unchanged except in the evaluation of the 
one body  Hamiltonian $h$. Due to the dependence on the density of the interaction, 
 $h$ gets \cite{MER.03} an extra term,  $\partial \Gamma$, which is usually referred to as 
the ''rearrangement potential'' and is given by
\begin{equation}
\partial \Gamma_{m m'} = \left< \frac{\partial H}{\partial \rho_{m'm}} 
\right>_T.
\end{equation}

 The   FTHFB solution gives us  the most probable shapes, quadrupole, hexadecupole, etc, 
as well as the most probable gap parameters and so on. At finite temperatures, however, 
we have statistical (or thermal) fluctuations around this solution.
In principle one could consider fluctuations in
the most relevant degrees of freedom. For nuclei, at high excitation energy, the most
important one is the quadrupole deformation, and we therefore shall consider only the fluctuations
in the quadrupole moment $\langle {\hat Q}_{20} \rangle $ in this paper. To generate the solutions 
with different shapes
we solve the grand canonical potential, Eq.~(\ref{gcp}), with an additional constraint on the
quadrupole moment, i.e., we minimize   \( \Omega = E - T S - \mu N - \lambda_{Q_{20}} q\). 
The Lagrange multiplier $\lambda_{Q_{20}}$ is adjusted in such a way that the thermal expectation value 
$\langle \hat{Q}_{20} \rangle = Tr(\hat{D}\hat{Q}_{20})$, has the required value $q$. 
 According to Landau \cite{LL.59} the probability $ P(q)$ to obtain a certain value $q$
of the deformation is characterized by the free energy
$F(q)= E(q) - T S(q)$ of the system with deformation $q$
\begin{equation}
   P(q)~\propto~e^{-F(q)/T}.
\label{probq}
\end{equation}
Using classical statistics, therefore, for the ensemble average of an observable $\hat{\cal O}$ 
one obtains  the expression
\begin{equation}
\label{averages}
 \overline{\cal O} =\frac{\int {\cal O}(q)\exp (-F(q)/T)dq}
{\int \exp (-F(q)/T)dq},
\end{equation}
where ${\cal O}(q)$ is the thermal expectation value of the
operator $ \hat{\cal O}$ calculated for the system with the deformation
$q$, and $dq$ is the volume element in deformation space.
 In our case the 
set \( q \)  corresponds to the quadrupole deformation \(q_{20}\),
thus \(dq = dq_{20}\), with metric equal to one. The limits in the thermal average integrals, 
see Eq.~(\ref{averages}), are chosen to span the full  $\beta_2$ region in which the probability 
of having one of these values, given by Eq.~(\ref{probq}), is not negligible. 
This covers both prolate and oblate regions.

High temperature calculations require  large configuration spaces. In order to
maintain the computational burden within reasonable limits we restrict ourselves
to axial symmetry. We are aware that for soft nuclei and/or high temperature, 
triaxiality may play an important role. 
 In the calculations we use an axially deformed harmonic oscillator (HO) basis with a 
size defined by the condition
 \begin{equation}
 2b_{\rho} n_{\rho} + b_z n_z < N_0,
 \end{equation}
where \( n_{\rho} \) and \( n_{z} \) are the HO axial quantum numbers,
\(b_{\rho}=q^{1/3} \) and \( b_{z}=q^{-2/3} \) with \( q=R_{z}/R_{\rho}, \)
the nuclear axis ratio. In our case we have used \( q=1.5 \) and \( N_{0}=15 \)
which allows for deformations big enough to reach the fission barrier and provides
room enough for the temperature induced excitations. However, as an additional check,
we have  also used \( N_{0}=17 \) for some selected calculations. 
Reflection asymmetry is allowed in the calculations, i.e. the nuclei may develop octupole
deformations.  

 In order to compare our Gogny force results 
  with the ones, more conventional and popular, of the schematic
Pairing plus Quadrupole model (PPQ), we have also performed calculations with this force. The
configuration space (the spherical oscillator shells $N = 5, 6$ for neutrons and  $N = 4, 5$
 for protons)
and the force parameters used are the one of Baranger-Kumar \cite{ppq}. 
The calculations have been performed in exactly the same way as in ref.~\cite{8}

\section*{Results}

We have performed FTHFB calculations with the D1S parameter set of the Gogny
force in several nuclei  to study the evolution of shell effects with temperature. 
 Nuclei with different ground state deformations  have been selected to illustrate their 
 different behavior.
As an example  of a nucleus with a strongly prolate deformed ground state  
we used the thoroughly studied \(^{164}\)Er. The soft \(^{152}\)Dy\(_{86}\) is
a transitional nucleus between the clearly spherical Dysprosium isotopes with 
$N \leq 84 $ and the well deformed ones with $N \geq 88$. This nucleus  was selected for
its rich shell structure and shape coexistence.  The heavier  \( ^{192} \)Hg has been chosen
due to its  oblate ground state shape. The maximum temperature studied
has been kept  below 3 MeV,  such that continuum contributions can be safely
disregarded~\cite{18,18b}.

 The thermal fluctuations are represented through averages calculated according
to Eq.~(\ref{averages}). The deviations around the mean values can be studied  
by the standard deviation value
\begin{equation}
\sigma ({\cal O}) = \sqrt{\; \overline{ {\hat {\cal O}}^2}  -
 \left[ \; {\overline{\hat {\cal O}}} \; \right]^2}.
\label{sig}
 \end{equation}
This quantity is  presented in some cases for further clarification of the results obtained.

 Before entering in the discussion of the shape and pairing phase transitions we 
will start by presenting first a general view.
  In Fig.~\ref{Fig:free} we present  the free energy, $F(\beta_2)$, and
the quantity \( P(\beta_2)~\propto~\exp (-F(\beta_2)/T) \)  versus the 
quadrupole deformation,  $\beta_2$, at different temperatures. \( P(\beta_2)\)
provides the weight of a given shape $\beta_2$ in the evaluation of thermal average
values. The results for $^{164}$Er are displayed in the left column,
for $^{152}$Dy in the middle one and for $^{192}$Hg in the right one.
 For the rare earth nuclei $^{164}$Er and  $^{152}$Dy, where calculations
within the two shells configuration space  mentioned above are feasible, we also present results 
with the PPQ model. Results with the Gogny force are displayed by continuous lines and
those with the PPQ force by dashed ones (thick lines represent $F(\beta_2)$ and thin
ones  \( P(\beta_2)\)).
The well depths are measured from the point with $\beta_2=0$ and \( P(\beta_2)\)  has
been normalized is such a way that the most probable deformation takes the value of unity.

 Let us discuss first the low temperature calculations
 ($T = 0.3 $ MeV ) where we can observe the intrinsic
shapes of the ground states. In the Gogny  calculations for $^{164}$Er there is a deep prolate 
minimum at $\beta_2 \approx 0.3$ and about 4.5 MeV higher an oblate one. With the PPQ model 
the same gross features are observed, though the minima are not so deep. The probability
distribution \( P(\beta_2)\), however, is similar in both calculations.
 For the nucleus $^{152}$Dy, in the Gogny case, the prolate minimum is at $\beta_2 \approx 0.15$ 
and a bit higher in energy the oblate one. The PPQ model, for this nucleus, provides a broad 
minimum around the spherical shape.  \( P(\beta_2)\), in contrast with $^{164}$Er,  looks quite
different in the Gogny case than in the PPQ one. In both nuclei the free energy surfaces are 
broader with the Gogny force than with the PPQ one.  For $^{192}$Hg, we find the minimum at 
an oblate deformation of $\beta_2 \approx -0.15$ and about 1.7 MeV higher a small prolate minimum.

At higher temperatures the expected disappearance of shell effects becomes clear, in particular
the vanishing of the barriers when several minima are available and the development of 
only one spherical minimum. Further finite temperature effects like the widening of the free 
energy curve and the more important role of fluctuations with increasing temperature appear
in the Gogny calculations but not in the PPQ ones. In the PPQ case the free energy surfaces,
with increasing temperatures, become flatter but not broader. They even become narrower! This 
unphysical effect has to do, obviously, with the size of the configuration space (two shells). 
As we can see already
at ($T = 0.3 $ MeV ), at large deformations  $F(\beta_2)$ increases very steeply because 
there are no orbitals with high-$j$ (deformation driving ) coming down. The mechanism to soften 
the free energy surface at high temperature  by enhancing the probability to occupy high-lying 
 orbitals (among them the high-$j$ ones) works only with large configuration spaces.
The anomalous behavior of \( P(\beta_2)\) in $^{152}$Dy, in the PPQ approach, at $T = 0.6 $ MeV
as compared with $T = 0.3 $ MeV  is due to the fact that at  $T = 0.6 $ the neutron pairing
gap vanishes and the ground state becomes slightly prolate. 

Figures~\ref{Fig:164ErAll},~\ref{Fig:152DyAll} and~\ref{Fig:192HgAll} show the 
detailed calculations for all three nuclei. These figures include both, the results
at the FTHFB level and with shape fluctuations calculated as described above.
Dashed lines and open symbols indicate the FTHFB results. Solid lines and filled symbols are used for
averaged, fluctuations including, calculations. Figs.~\ref{Fig:152DyAll} 
and~\ref{Fig:192HgAll} show  only Gogny results.

\subsection{The nucleus $^{164}Er$}

In Fig.~\ref{Fig:164ErAll} we  display the results of the calculations for the 
nucleus \( ^{164} \)Er with the Gogny force and with the PPQ model Hamiltonian.
In panel (a) we show the selfconsistent FTHFB (i.e., calculated with the solution 
of Eq.~\ref{hfbeq}) and the  averaged results (i.e., calculated according to 
Eq.~\ref{averages})  for the $\beta_2, \beta_4$ and $\beta_6$ deformation parameters as a 
function of the temperature with the Gogny 
interaction. Let us first discuss  the deformation parameter $\beta_2$. For temperatures 
$0<T<1.0$ MeV, both predictions behave similarly,  as 
one would expect for a nucleus with a well pronounced minimum. 
For temperatures $1.0<T<2.0$ MeV, the FTHFB $\beta_2$-values  decrease rather smoothly
while the averaged ones undergo a strong  reduction. For $T>2.0$ 
MeV the selfconsistent values decrease very steeply and collapse, finally, to zero deformation 
at $T=2.7$ MeV. The averaged values, on the contrary, change tendency decreasing very smoothly
in such a way that an almost  constant value of  $\beta_2$ is eventually obtained. The behavior of
 $\beta_4$  and
$\beta_6$ is similar to the one of $\beta_2$ though not that spectacular. The same plot
for $\beta_2$ but with the PPQ interaction is represented in panel (b). Quantitatively the
main differences with the Gogny results are the faster collapse of the selfconsistent value, 
at $T\approx 1.8$ MeV, and the reduction of the temperature interval where the averaged values
are smaller than the selfconsistent ones. Looking at the probability distribution in 
Fig.~\ref{Fig:free} one can  easily understand these differences. The temperature value at which
the {\em mean field} (mf) deformation parameter collapses, which we will denote $T^{mf}_{c}$,
has often  been  used in earlier
mean field studies to signal a shape phase transition. It is obvious
from panels (a) and (b) that in theories {\em beyond mean field} things look quite different and that 
definition of  the critical temperature must be carefully considered.
The big difference in $T^{mf}_{c}$ as predicted by effective forces, like the Gogny force, 
and the PPQ  is also known from calculations with  Skyrme forces~\cite{BQ.74a,BQ.74b} and
the relativistic mean field approximation~\cite{AT.00,10b}.

The standard deviation in the deformation parameter $\beta_2$, $\sigma(\beta_2)$, calculated according 
to Eq.~\ref{sig}, is presented in panel (c). One can distinguish three  well defined zones~: in the first 
one at low temperature, when pairing is still strong, the  deformation is kept almost constant and 
fluctuations raise slowly. The Gogny and PPQ calculations predict about the same equilibrium shape 
in this zone.  For temperatures higher than the one corresponding to the  pairing collapse, 
 $\sigma(\beta_2)$ increases rapidly up to a maximum value, remaining more or less at this value at higher
 temperatures. 
 This step behavior is characteristic of a shape phase transition region. 
In fact, one could  define the shape transition temperature as the one at which $\sigma(\beta_2)$
 has a maximum. With this criterion one obtains $T=1.4$ MeV for the PPQ result 
and $T=1.7$ MeV in the Gogny case.  Note the large difference in  $\sigma(\beta_2)$ between the 
Gogny and PPQ results at high temperature. In this comparison, however, one must keep in mind that
the PPQ model hamiltonian is restricted to a configuration space 
of two oscillator shells which strongly constraints the ability to produce
fluctuations. This is clearly seen in Fig.~\ref{Fig:free}~: the PPQ results rapidly develop a 
narrow parabolic shape with increasing temperature. This lack of 
fluctuations was already identified as partially responsible for the low multiplicity seen in the
collective E2  quasicontinuum spectra in gamma decay calculations when
compared to experiment~\cite{16,16b}.

An additional  confirmation of the importance of the fluctuations in calculations
with  the Gogny interaction as 
compared to the PPQ case is provided by the different behavior of the 
averaged deformation parameters with respect to the mean field  within the same model, see
panels a) and b).   The value of the average $\beta_2$ parameter {\em at high
temperatures} illustrates the deviation of the free energy surface from a parabolic behavior.
A value close to zero is expected for a parabola, e.g., in the PPQ case,  while a larger one,
as in the Gogny case, indicates the softness of the prolate side as compared with the oblate one. 
One should nevertheless keep in mind  that only axially symmetric deformed shapes are allowed 
in the calculations.

In panel d) of Fig.~\ref{Fig:164ErAll}, the proton and neutron pairing energies are displayed 
for the Gogny force. Up to $T \approx 0.2 $ MeV the pairing energies are rather constant but
for higher $T$ values they decrease in absolute value very fast up to $T = 0.7$ MeV where they 
vanish. Thermal shape fluctuations, as expected in the low temperature regime, have little effect on 
the pairing correlations.  Pairing fluctuations which would be more relevant \cite{ERI.85} are not
considered in this work.

It is interesting to take a look at the internal excitation energy, $E^*$, and 
also analyze it through the behavior of its derivative, the specific 
heat $C_V(T)=\partial E^*/\partial T$, since the appearance of peaks in 
this quantity is customarily interpreted as a signature in the search for phase transitions.  

The evolution of  $E^*$  with temperature for both the FTHFB,  $E^*$,  and the average,
$\overline{E^*}$, calculations, is presented in panels e) (Gogny force) and  f) (PPQ  force)
of Fig.~\ref{Fig:164ErAll}.  With  the Gogny force and in the low temperature regime, 
we can see the pairing collapse which is visible as a change in the slope of $E^*$. 
  At higher temperatures  a fairly quadratic behavior  is observed 
in the excitation energy,  which is slightly modified when the transition to the spherical
phase takes place at high temperature around $T \sim 2.7$ MeV. There, a weaker 
change in the slope, hardly seen in the scale of the plot, is found.
The change is more abrupt in $E^*$ than in $\overline{E^*}$, 
again as expected in the picture of a thermally faded
transition. The same facts are observed in the PPQ plot, where one can additionally observe
that at high temperature the energy behaves  more linear than quadratic as a function of T. 

The different behavior of  $E^*$ and  $\overline{E^*}$ is also interesting~:
a) At temperatures below 0.8 MeV, both energy values coincide, b) between 0.8 MeV and the corresponding
$T^{mf}_{c}$ (around 2.7 MeV for Gogny and 1.8 MeV for PPQ),   $E^*$ is always below
 $\overline{E^*}$, c) at the critical temperature both energies do coincide and d)
 at higher temperatures  they are rather similar.
This behavior has a simple explanation if one considers the entropy as a function
of the deformation  at fixed temperature, see Fig.~\ref{Fig:Entro}, and the fact that in general
 $F \leq \overline{F} $ \footnote{ The equal sign is valid only in the case that the free energy 
is a parabolic function of the deformation.}, with $F$ the selfconsistent FTHFB
free energy at the given $T$. At low temperatures ($\leq 0.8$ MeV) and at temperatures 
above $T^{mf}_{c}$, the entropy is almost shape independent, i.e., $\overline{S}  
\approx S $ (with $S$ the  selfconsistent FTHFB entropy at the given $T$), 
and the free  energy behaves like a parabola, $ \overline{F}\approx F $, 
consequently  $ \overline{E^*} \approx E^*$. At temperatures $0.8 \leq T \leq T^{mf}_{c} $,  
 $F < \overline{F}$ and $S < \overline{S}$, see Figs.~\ref{Fig:free} and \ref{Fig:Entro}, 
 and consequently, since $\overline{F}= \overline{E} - T \overline{S}$,  $E^* < \overline{E^*}$.

 The change in slope in $\overline{E^*}$ as compared with $E^*$ indicates that the corresponding
specific heats, $ \partial \overline{E^*} / \partial T$ and $\partial E^* / \partial T$, 
will be rather different.
This can be seen in Fig.~\ref{Fig:Cv} where we observe two peaks in the selfconsistent 
results, both in the PPQ and in the Gogny calculations 
\footnote{The results displayed in this figure have 
been calculated by evaluating numerically the derivative of the energy. In ref.~\cite{EMR.00} 
the expression $ C_V = T \partial S/\partial T $ was
used instead and $S $ was calculated analytically in the mean field approach.}. In between
we find the typical linear behavior for a Fermi gas (\( C_V = 2aT \)).
The low temperature peak is associated with the superfluid to normal transition and the high 
temperature one with the deformed to spherical shape transition. 
 The low temperature peak remains nearly unaffected by the inclusion of
fluctuations. At the mean field level this transition takes place at a temperature low enough 
such that shape fluctuations are irrelevant. 

Comparing the FTHFB  results with the  PPQ force with those obtained with the Gogny  interaction
for the shape transition,  we find  again the different temperature predictions. Using as  critical 
temperature, $T_{c}$, the temperature where $C_V$ changes curvature, the same values are obtained as when
the $\beta_2=0$ rule was used (1.8 MeV for PPQ {\sl vs.} 2.7 MeV for Gogny). However, when 
fluctuations are taken into account, the difference in the predicted  $T_{c}$ by Gogny and PPQ 
gets smaller and the sharpness of the peaks  reduced, indicating a less abrupt transition
 as is expected in a  mesoscopic system.
This is more evident in the Gogny results, where the peak becomes a broad bump, providing
another clue of the greater importance of fluctuations in the Gogny case. By contrast, the
PPQ peak, although broader than in the mean field case, is still sharp. Furthermore the PPQ 
specific heat
levels off, showing how the limited configuration space available is a clear disadvantage of this model.   
If we now look for the changes in curvature we find $T_{c}=1.4$ MeV in the PPQ and $T_{c}=1.7$ MeV
with the Gogny interaction. It is interesting to notice that these values agree very well with the 
ones obtained looking at $\sigma(\beta_2)$.


\subsection{The nucleus $^{152}Dy$}

  As we have seen in Fig.~\ref{Fig:free},  $^{152}Dy$ displays a potential energy surface 
with energetically close prolate and oblate minima. It could illustrate a nucleus with shape coexistence, 
that means, already at temperatures near to zero there is a finite probability of populating more 
than one minimum.  For this nucleus we will not perform a discussion as exhaustive as for $^{164}Er$ 
but we will consider the most relevant facts.

  In Fig.~\ref{Fig:152DyAll} the results of the calculations for the Gogny force are displayed.
In panel a) the mean field and the averaged values of the pairing energies are plotted. The pairing 
energy of the neutron (proton) system collapses at $T \approx 0.5$ ($T \approx 1.0$) MeV. The 
 averaged values, as expected, almost coincide with the mean field ones. 
In panel b) the $\beta$-deformation parameters are
shown. As we can observe in the behavior of the  $\beta_2$ parameter the effect of the shape fluctuations
in this case is already noticeable at very small temperatures. This is due to the 
fact that, in the ground state  $^{152}Dy$ is a much less deformed nucleus 
than  \( ^{164}\)Er and that the energy difference between the oblate and prolate  
minimum is small, amounting to only $\sim 0.55$ MeV for $T<0.5$ MeV. Hence the 
averaging formula assigns finite  weights to the oblate side already at small $T$'s 
causing the observed steep decrease in the average $\beta_2$ value. Although the FTHFB, 
searching for the strict minimum, provides a deformed ground state for $^{152}$Dy, we see how even 
at the  lowest temperatures the average deformation is very small, in agreement  with the
experimental data. At the high temperature limit we observe that above 1.4 MeV, the average deformation
stays rather constant, or slightly increases, up to 0.04. This anomalous behavior is due to
the fact that the superdeformation driving orbitals are being occupied at this temperature range.
The  \( \beta _{4} \) and \( \beta _{6} \) deformation parameters follow closely the 
 $\beta_2$ behavior. In particular, in the FTHFB description they become zero 
 at the same temperature as $\beta_2$. 
 
In panel c) the excitation energy is depicted. In the FTHFB approach the changes in slope 
 at temperatures of 0.5, 1.0 and 1.4 MeV are due to the  neutron and proton pairing collapse and
to the shape  transition. The behavior of $\overline{E^*}$ can be understood in terms of the entropy
plots as explained for the $^{164}$Er nucleus. The changes in slope in the energy plots are magnified 
in the specific heat versus temperature plot depicted in panel d).  In the mean field approach we find 
a broad composite peak, corresponding to the proton and neutron pairing collapse, showing substructures
around $T\approx 0.5$ and 1.0 MeV. Furthermore, one sees a second peak at 1.4 MeV corresponding
to the FTHFB shape transition.
 If shape fluctuations are included in the calculations we obtain only one broad peak. The small peak
at 1.4 MeV, however, is not there anymore indicating that it has been shifted to the pairing 
transition bump or simply washed out.  
  In fact, the almost identical broad energy mean field peak and the single one obtained with
fluctuations could be seen as a clue that the shape transition in this soft nucleus is
inexistent, since there is no higher temperature peak but only a small modification 
of the "pairing" one. 
To check  this hypothesis we have also performed calculations with the particle-particle 
channel of the Gogny force set to zero, i.e. plain FTHF. In this way we obtain $C_V$ curves without the  
pairing transition peaks. The results are plotted, superimposed to the standard FTHFB
calculations, using square symbols in panel d) of Fig.~\ref{Fig:152DyAll}.
Again, open squares are for the FTHF calculations and full ones for averaged ones. As it was expected, 
the FTHF curves for  $C_V$  show no peak for the pairing transition and the only peak present is the 
one corresponding to the shape transition at 1.4 MeV, which coincides with the one obtained with the
full FTHFB calculations, since at these temperatures pairing is already zero. The thermal averaging results 
show a broad shoulder approximately in the same temperature region in which the pairing transition 
was located. This rather soft bump is a clear indication of the above mentioned situation, i.e., at small
temperatures no clearly predominant minimum exists and at high temperatures the nucleus does not 
become exactly spherical.

\subsection{The nucleus $^{192}$Hg}

As we have seen in Fig.~\ref{Fig:free} at very small temperatures this nucleus presents an
oblate deformed ground state and about 1.7 MeV above a prolate minimum. 
 The results for the nucleus $^{192}$Hg are shown in Fig.~\ref{Fig:192HgAll}. The pairing energies
are displayed in panel a). The proximity of the $Z=82$ shell closure causes the vanishing of the
proton pairing energies for all temperatures. The neutron system, on the contrary, has a large
pairing energy at $T=0$ MeV, which vanishes at $T=0.8$ MeV. As before, the shape fluctuations have
almost no effect on the pairing energies.  

   In panel b) the behavior of the deformation parameters with increasing temperature is plotted.
 For $^{192}$Hg, in the mean field approximation, we obtain an oblate ground state deformation of 
$\beta_2 = -0.135 $ which gets more oblate for increasing temperatures as the pairing energies 
go to zero. For larger $T$ values the deformation decreases and around $T \approx  1.4$ MeV the
nucleus becomes spherical. As before, the effect of the shape fluctuations is mainly characterized by the
the prolate-oblate ground state energy difference, see Fig.~\ref{Fig:free}, which in this case  
amounts to 1.7 MeV at zero temperature and only above $T=0.5$ MeV start to  diminish. Around this 
temperature
the deformation gets smaller and around 1.4 MeV the average deformation is zero.
  Interestingly, the average $\beta_2$ and $\beta_4$
deformation parameters become positive in the limit of high temperatures due to the fact that at these
temperatures the $F(T)$ curves are softer in the prolate than in the oblate side.

  In panel c) the excitation energies in both approximations, in the mean field and with shape
fluctuations,
are plotted versus the temperature. In the mean field calculations we find slope changes at
$T=0.8$ MeV and $T=1.4$ MeV associated with the pairing collapse and the oblate-spherical
shape transition. The inclusion of shape fluctuations affects mainly the region between 
$0.8 \leq T \leq 1.4$ MeV. The general behavior of both curves can be easily
understood in the same terms as for $^{164}$Er. In panel d) the corresponding specific heats are
represented. In the mean field approximation the expected peaks are clearly visible. The inclusion
of the shape fluctuations produces a single, broader bump extending above the critical temperature for
the pairing collapse. As in the  $^{152}$Dy case this peak might be a superposition of the pairing and
the shape transition peaks. To isolate the shape transition peak we have performed again calculations
with the particle-particle channel of the Gogny force set to zero. The results of the calculations,
without (empty squares) and with (filled squares) shape fluctuations, are represented in the same panel.
In both approximations the results above $T=0.9$ MeV are obviously the same as before. Below this
temperature and in the mean field approximation, as expected, the pairing peak is gone. However, in
the calculations with the shape fluctuations we find a broad peak extending from $T=0.4$ MeV up to
$T=1.0$ MeV with a change in curvature around $T=0.9$ MeV. Looking at the standard deviation
$\sigma(\beta_2)$ of this nucleus (not shown here) we find a maximum at $T=0.9$ MeV, an additional
indication of the shape transition.

\subsection{Level densities and nuclear radii}

 Level densities, $\rho(E^*)$, can be microscopically evaluated in the saddle point 
approximation,  see for example eq.~(2B-14) of ref.~\cite{BM.75}.  
In Fig.~\ref{Fig:lden} the total level densities for the three nuclei under study are
displayed against the excitation energy in the mean field approximation and on average, i.e., 
with the inclusion of shape fluctuations.
 In both cases we observe the overall expected exponential dependence and the 
well known abnormal behavior at very small excitation energies. 

For  $^{164}$Er and up to 10 MeV excitation energy ($T \approx 0.8$ MeV) we find a good agreement
 between both 
predictions. Then, up to 70 MeV ($T \approx 2.3$ MeV), we observe an increase in the 
level density in the average description as compared with the mean field one. In particular,
around 20 to 30 MeV excitation energy, the average prescription provides almost two orders of 
magnitude larger densities than  the mean field one. This behavior can be easily understood 
looking at Figs.~\ref{Fig:free} and \ref{Fig:Entro} and taking into account that the level density 
is proportional to the exponential of the entropy. At low ($T < 0.7 $ MeV) and high excitation 
energies ($T > 2.5 $ MeV) the entropy is rather shape independent, that means, the average 
value of the level density is very close to the one in the selfconsistent minimum.  
Consider now  $T=1.4$ MeV. In this case the selfconsistent minimum is prolate ($\beta_2 \approx 0.3$),
and the entropy at this shape and $T$ is smaller than for all the other shapes at this 
temperature. That means, since  $\rho(\beta_2) \propto e^{S(\beta_2)}$, the average level 
density will  always be larger than the selfconsistent one.  Similar arguments apply to understand
the behavior of the level densities of  $^{152}$Dy and  $^{192}$Hg. The fact that in these nuclei
we do not find a larger difference between both descriptions is obviously due to the smoother
behavior of the entropy with the deformation at the relevant temperatures. 

In Fig.~\ref{Fig:msr} the root mean squared (rms) radii  of the three nuclei are plotted versus the 
temperature. In general we find that the rms radii are 
rather constant up to a given temperature, 2 MeV for $^{164}$Er and 1 MeV for $^{152}$Dy and  
$^{192}$Hg, and that in this temperature range the average values are 
rather similar to the FTHFB ones. From this temperature on the average values are larger
than the FTHFB ones due to the fact that, at these temperatures, the probability for a given 
shape peaks at the spherical shape and that for a given volume the spherical shape  corresponds 
to the one with the smallest rms radii. That means, fluctuations around the spherical minimum provide
always larger rms radii.
We also observe, at the highest temperatures, the expected  increase of the rms radii.

\section*{Discussion and conclusions}

We have seen in the previous section that, in calculating average properties, the behavior of the
entropy with the deformation parameter $\beta_2$ plays a major role.  This behavior 
is by itself, indeed, quite interesting. The general behavior, see Fig.~\ref{Fig:Entro}, is 
the following~: 
 In the high $T$ limit  where the temperature effects dominate, we find, as expected,  small 
shape dependence. At very low $T$, where the temperature effects are very small,  we observe that to increase the 
entropy by 5 units we have to increase $T$ by 0.5 MeV and that the entropy is rather independent 
of the shape of the nucleus. Of course, in this region where  pairing correlations 
are present it is difficult to make more precise statements.
However, at moderate temperatures, which are however high enough  to allow for significant
quasiparticle occupation numbers but not too high in order that shell effects are still present, 
one can find a large dependence of the entropy on the nuclear shape. In this region spherical shapes, 
as expected, have larger entropy than axially deformed ones. Since the maxima of the entropy are 
associated with  the minima of the grand potential a correspondence between Fig.~\ref{Fig:free} and 
Fig.~\ref{Fig:Entro}  does not necessarily exist.

One of the main outcomes of our research is the finding that shape fluctuations have a large
effect on the description of  shape transitions. In fact, the {\em mean field approach} (FTHFB) with 
effective forces (like Skyrme, the Gogny force or the relativistic approaches), provides the view of 
a sharp shape transition at a relatively high critical temperature ($T_c \approx 2.7$ MeV for
$^{164}$Er). On the other hand, Strutinsky calculations or schematic models (like the Pairing plus 
Quadrupole) provide also a sharp transition though at a much lower critical temperature 
($T_c \approx 1.7$ MeV  for $^{164}$Er). It has been argued \cite{GML.00} that the different 
predictions for the critical temperature are due to the small effective mass obtained in the mean field
approach with effective forces ($m^*/m \approx 0.7$, with $m$ the nucleon bare mass ) as compared to the 
Strutinsky or the PPQ model ($m^*/m \approx 1.0$).  This argument is obviously restricted to the
mean field approach.  In theories beyond mean field it does not apply anymore because with increasing
correlations the effective mass eventually becomes the bare mass. In calculations at finite temperatures two 
kinds of correlations have to be considered, on one hand the quantum ones and on the other 
 the statistical or classical ones. Their relevance depends obviously on the excitation energy 
 (or temperature), at low
$T$'s the former are very important and at high $T$'s, the latter ones. At the temperatures 
where the shape transition is predicted to take place in the mean field approach,  the probability 
of having a shape different to the selfconsistent one is very large.  Therefore,
it is obvious that, first, shape
fluctuations must be included and second that the characterization of the shape transition
must be considered more carefully. In the mean field approach a criterion for shape transition
is just to look at the temperature at which the nucleus becomes spherical or alternatively to look
for a peak at the specific heat as a function of the temperature. In theories beyond mean field, 
usually the second one is used because the average deformation can become very small but not zero.
As we have seen, the inclusion of shape fluctuations provides a specific heat rather different
from the mean field one,  because a) it is not a sharp peak what we obtain but a rather broad bump (this is
consistent with the fact that it is a very small system) and b) it appears at temperatures much lower
than the ones predicted by the mean field approximation in agreement with the Strutinsky calculations. 
It is also interesting to notice that the predictions based on the specific heat analysis coincide
with the ones of the standard deviation $\sigma(\beta_2)$.

Of course one could ask about the effect of considering quantum correlations in our predictions.
Canosa, Rossignoli and Ring \cite{CRR.99} have shown in model calculations based  
on the static path plus random phase approximation that at finite temperature quantum effects 
are observable dependent.  In particular, they find that the specific heat remains practically unaffected 
when quantum correlations are taken into account.  
One could conclude therefore that the prediction of the shape transition by the inclusion of
thermal shape fluctuations with effective forces is reliable.

   In conclusion, we have performed extensive calculations with the Gogny force and a large 
configuration space for three representative nuclei in the FTHFB framework. We have further studied
 the effect of thermal shape fluctuations and found that they strongly affect, among
others, the traditional shape transition "view" of the FTHFB approach. They do it in two aspects,
first the critical temperature for the transition is very much lowered and second, the specific heat peaks
are not sharp but rather broad. Besides this, the peaks showing up in FTHFB calculations of the specific
heat in soft nuclei, like  $^{152}$Dy, are (almost) completely washed out when thermal shape
fluctuations are taken into account indicating the absence of any shape transition. However,
in strongly deformed nuclei, like $^{164}$Er,  the shape fluctuations reconfirm the presence of
a shape transition though of a different character. We also find a strong enhancement in the level
density in the presence of a shape transition. The superfluid to normal phase transition is not affected
by the inclusion of  shape fluctuations.

\begin{acknowledgments}
This work has been supported in part by DGI, Ministerio de Ciencia y
Tecnolog\'\i a, Spain, under Project BFM2001-0184.
\end{acknowledgments}

\bibliography{fluc}


\begin{widetext}

\begin{figure}

{\centering {\includegraphics[width=0.85\textwidth]{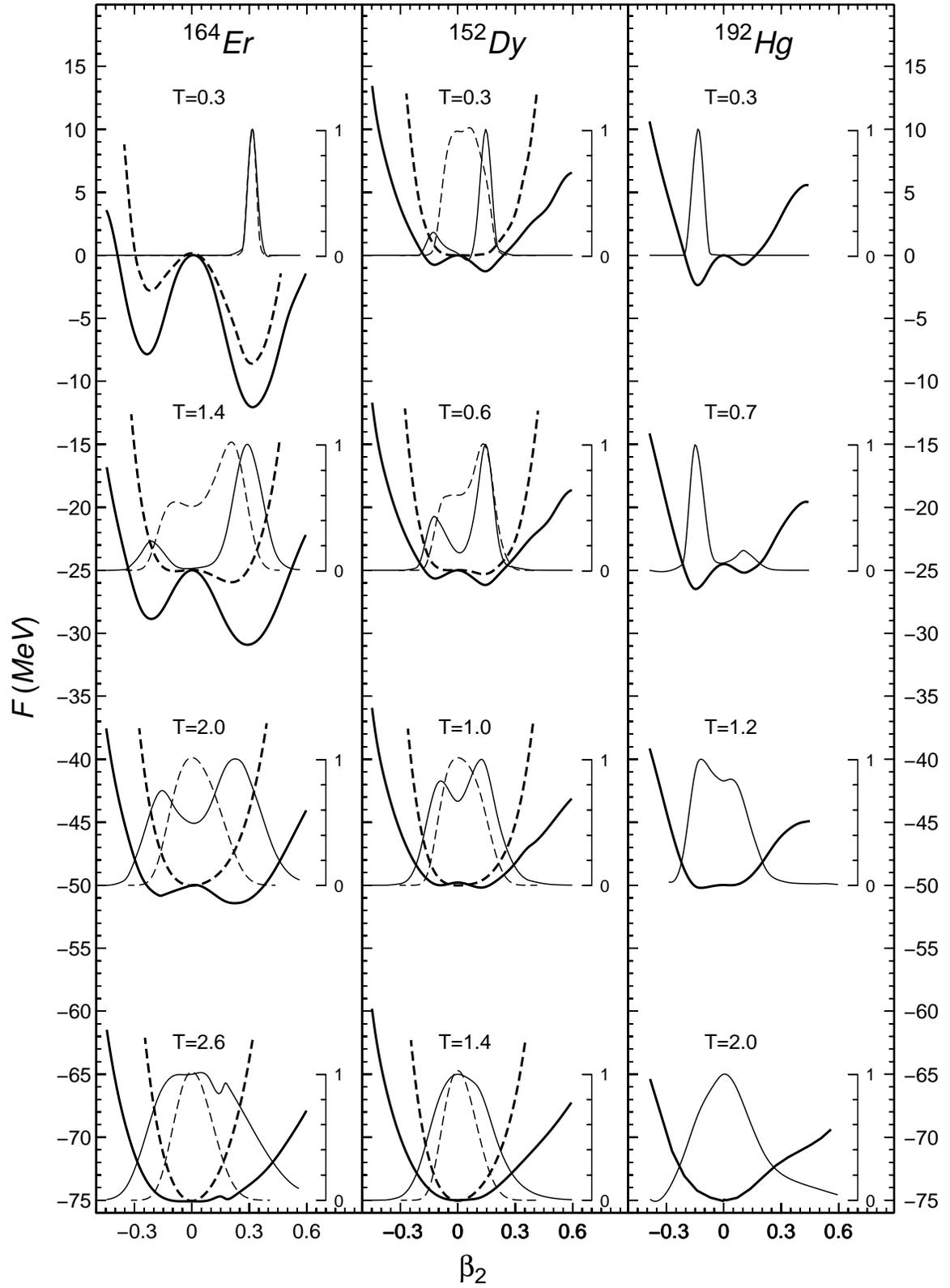}} \par}

\caption{Free energy curves  for $^{164}$Er, $^{152}$Dy  and $^{192}$Hg
 calculated with the Gogny (thick solid lines) and PPQ (thick dashed lines) interactions
at several temperatures as a function of the quadrupole deformation 
parameter $\beta_2$. The probabilities $P(\beta_2)$ for a given shape $\beta_2$, with the Gogny
(thin solid lines) and the PPQ (thin dashed lines) interaction. The scale for $P(\beta_2)$
is given in the inset. 
}

\label{Fig:free}

\end{figure}

\begin{figure}

{\centering {\includegraphics[width=0.75\textwidth]{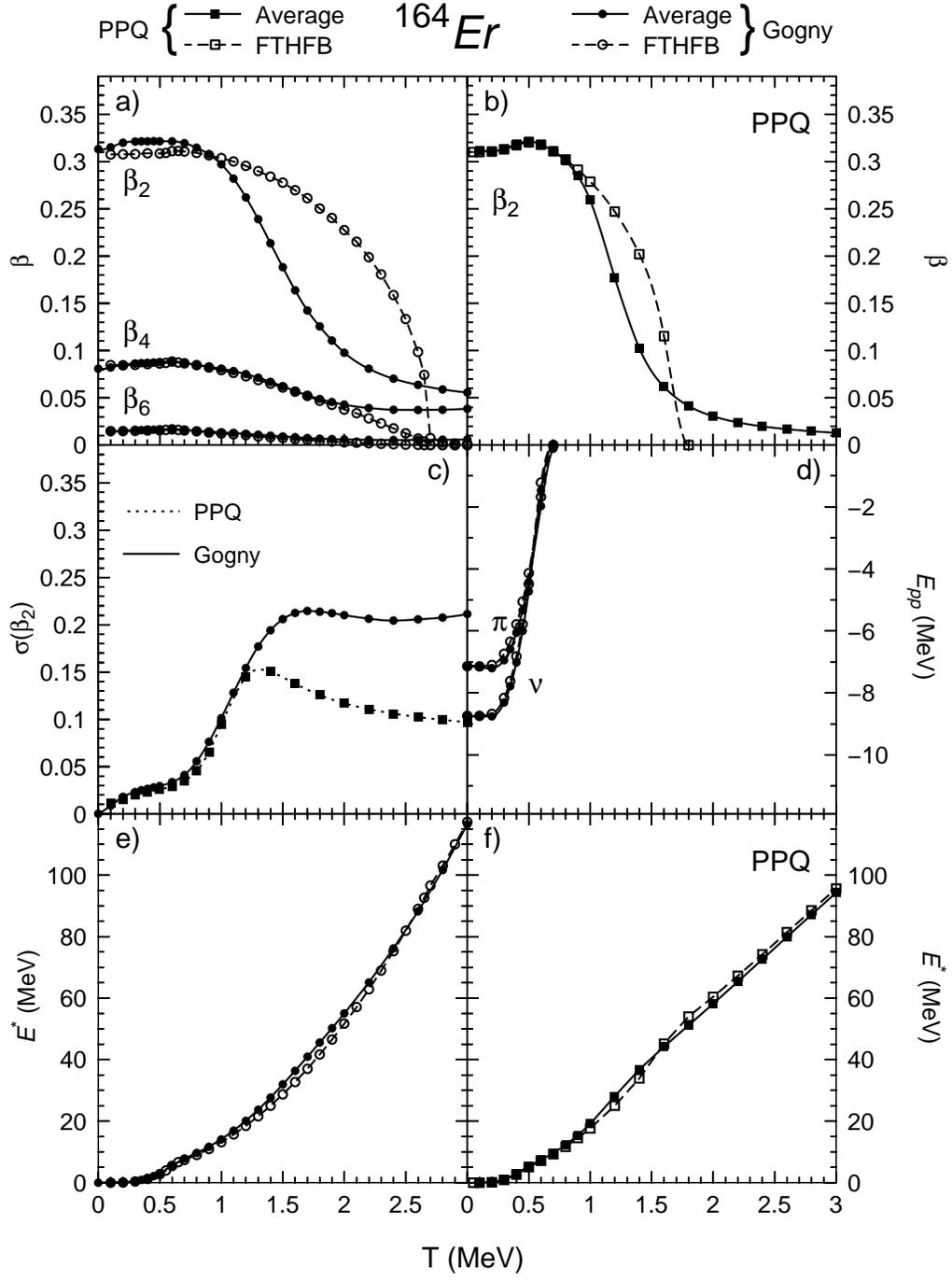}} \par}

\caption{
 Results for \( ^{164}\)Er for several observables as a function of the temperature.
In all panels except in (d), solid lines and filled symbols are average  values calculated
according to Eq.~\ref{averages}, dashed lines and open symbols are for selfconsistent (FTHFB)
 results. Circles are for Gogny and squares for PPQ. Shown are the
 $\beta_2$ deformation parameter with Gogny (panel a) and PPQ (panel b) forces, the standard deviation $\sigma (\beta_2)$ with Gogny and PPQ forces (panel c), the pairing 
 energies with the Gogny force (panel d) and the excitation energy with the Gogny  (panel e) and
 the PPQ force (panel f).}
\label{Fig:164ErAll}
\end{figure}

\begin{figure}
{\centering {\includegraphics[width=1.0\textwidth]{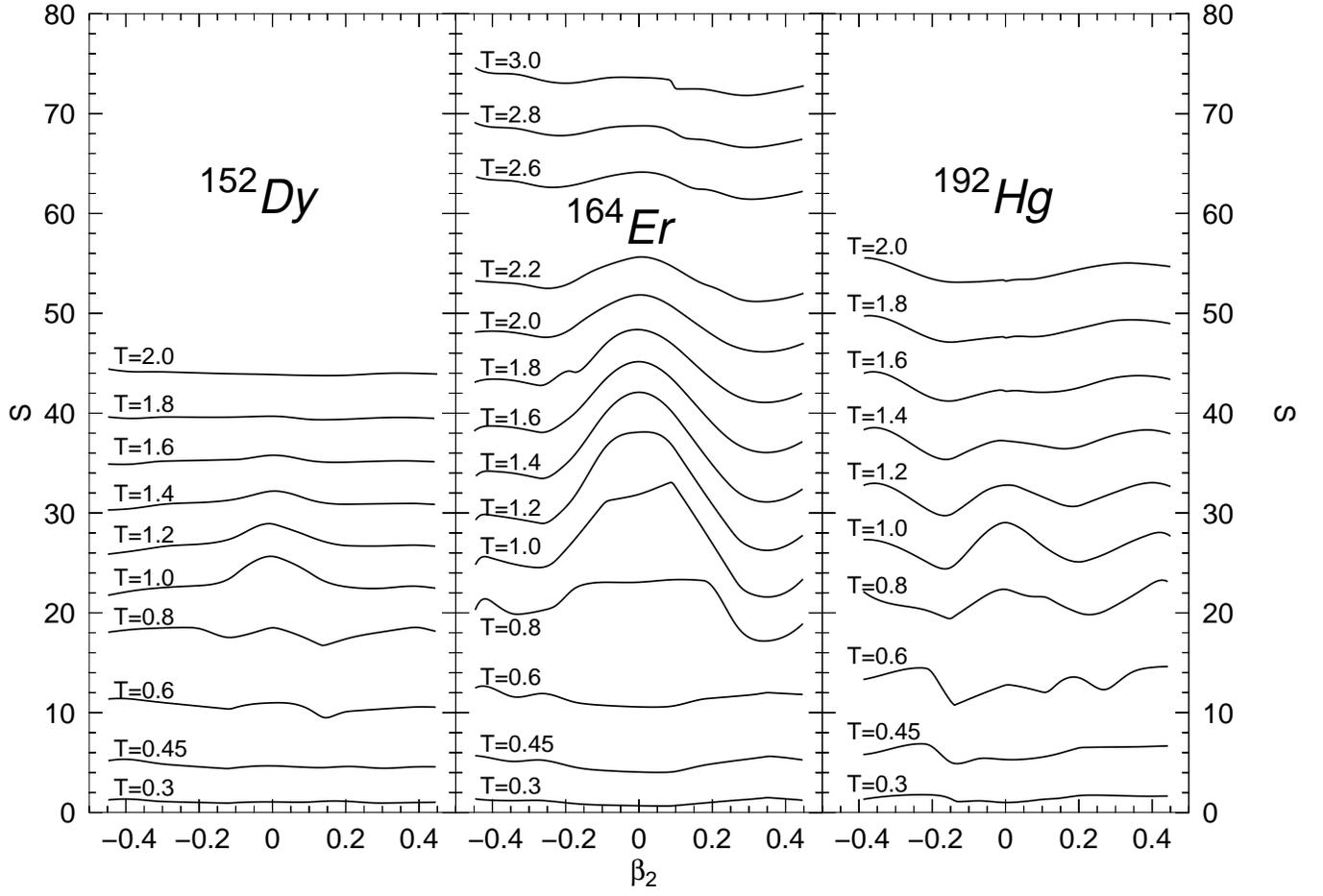}} \par}
\caption{ The entropy at fixed temperatures as a function of the deformation parameter
$\beta_2$.}
\label{Fig:Entro}
\end{figure}


\begin{figure}
{\centering {\includegraphics[width=0.9\textwidth]{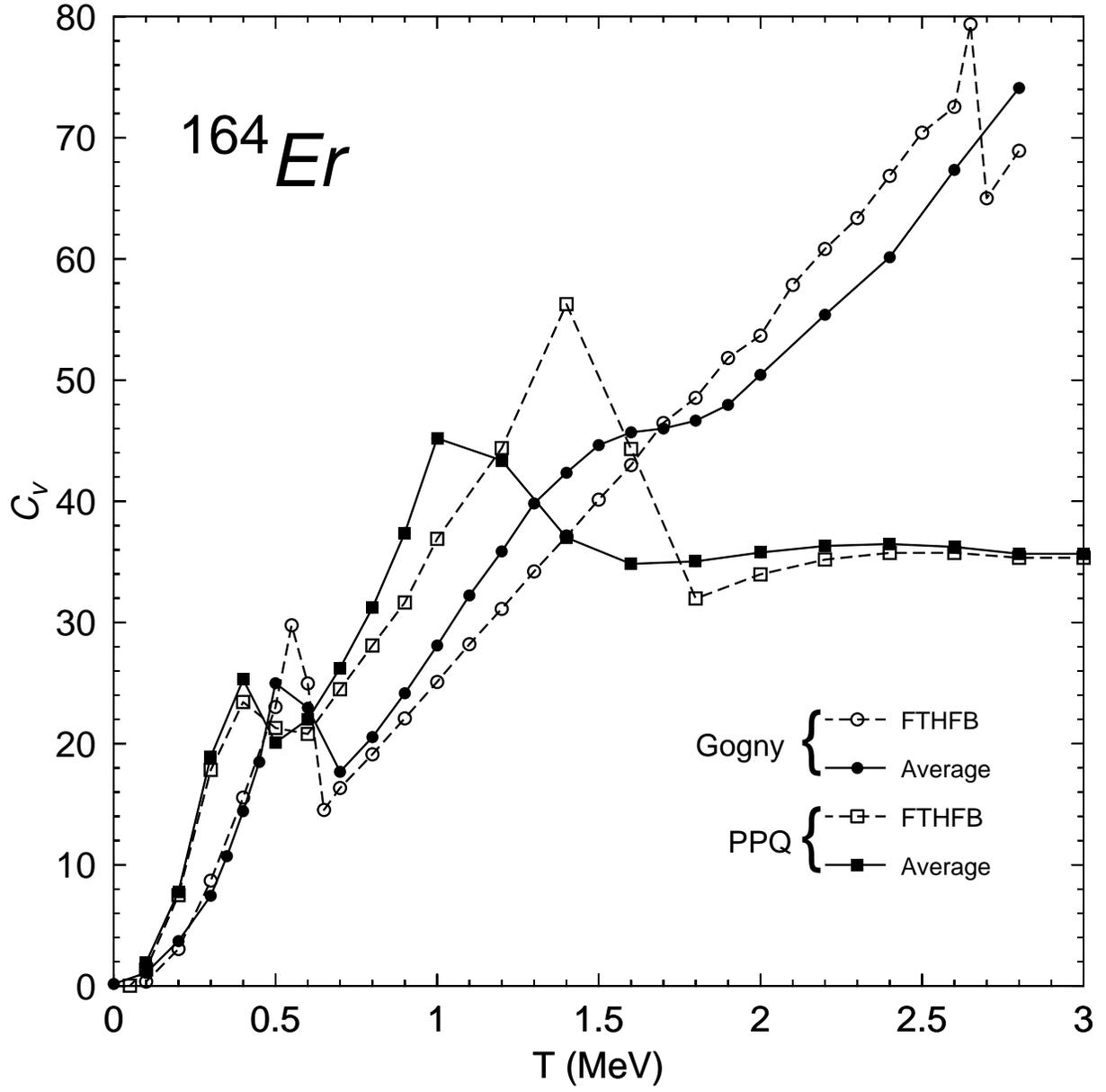}} \par}
\caption{ The specific heat for the nucleus \( ^{164}\)Er with  the Gogny  (circles)
and the PPQ (squares) interactions, in the FTHFB approximation (empty symbols) and the
averages with shape fluctuations (filled symbols). }
\label{Fig:Cv}
\end{figure}


\begin{figure}

{\centering {\includegraphics[width=0.9\textwidth]{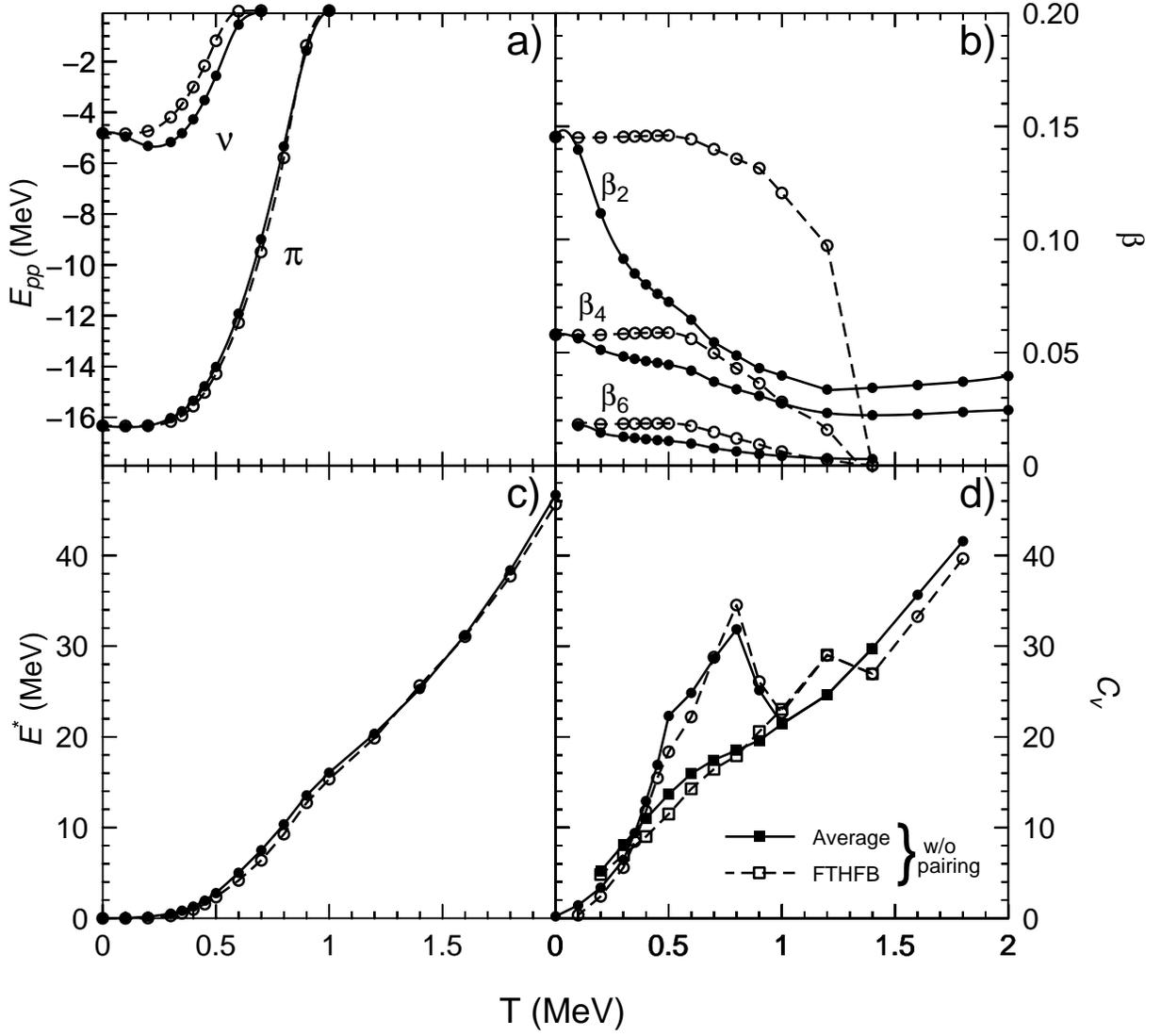}} \par}

\caption{ Results for \protect\( ^{152}\protect \)Dy with the Gogny force versus the
temperature. Average values are presented as solid lines and filled circles. 
Selfconsistent FTHFB results as dashed lines and open circles. 
Upper row, panel (a), pairing energies for  protons $\pi$ and neutrons $\nu$. 
Panel (b), deformation parameters. Lower row, panel (c), excitation energies. Panel   
(d) specific heat calculated as defined in the text. Square symbols correspond to results 
using the Gogny force with pairing set to zero. In the temperature range from 1 to 1.5 MeV
squares and circles are superimposed since above 1 MeV the solutions with and without pairing
are the same.  
}
 
\label{Fig:152DyAll}
\end{figure}


\begin{figure}

{\centering {\includegraphics{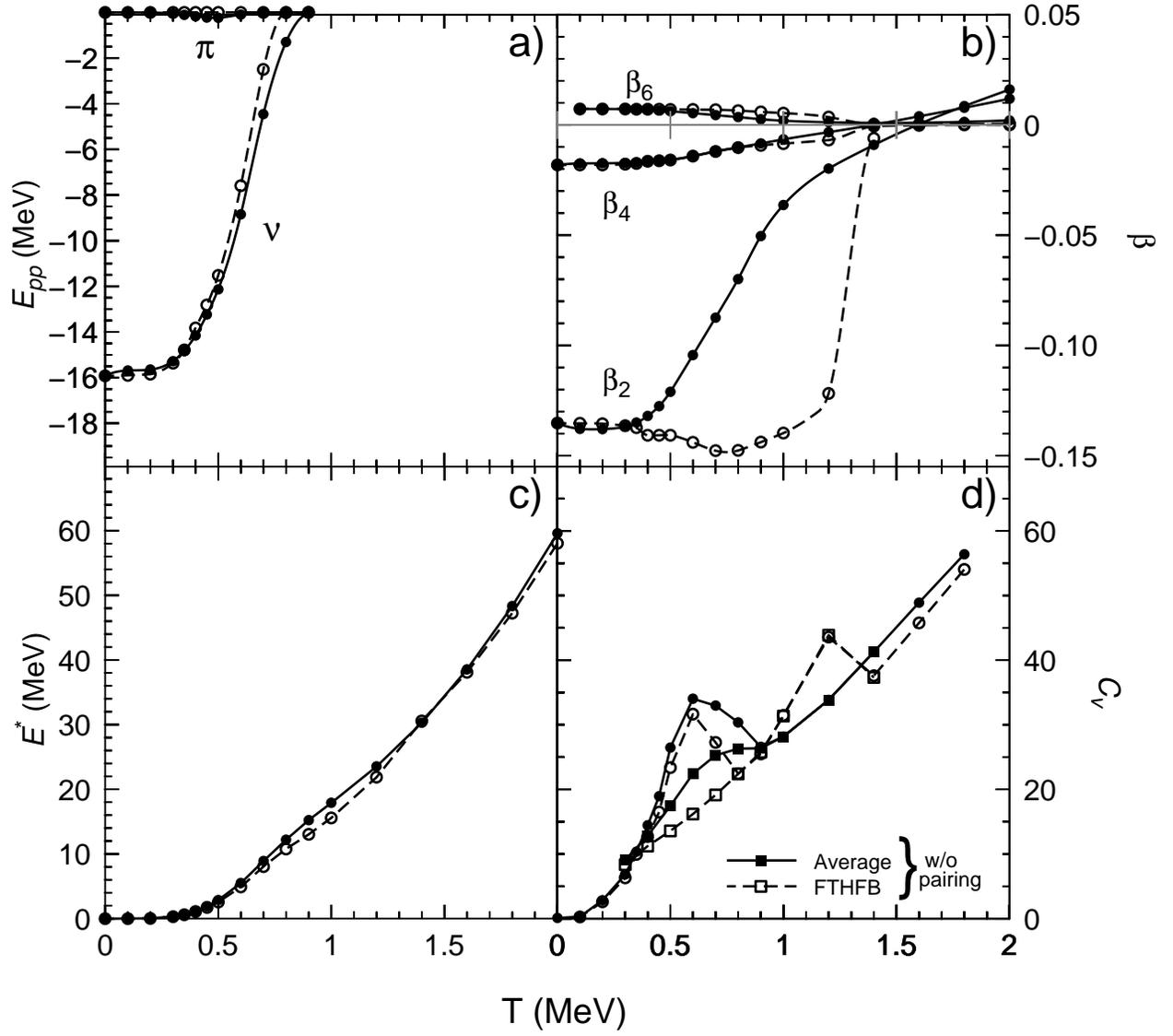}} \par}

\caption{ Same than Fig. \ref{Fig:152DyAll} for the nucleus
\protect\( ^{192}\protect \)Hg. In panel (d) the superposition of results with
and without pairing starts at T=0.9 MeV.
}
\label{Fig:192HgAll}
\end{figure}

\begin{figure}
{\centering {\includegraphics[width=0.8\textwidth]{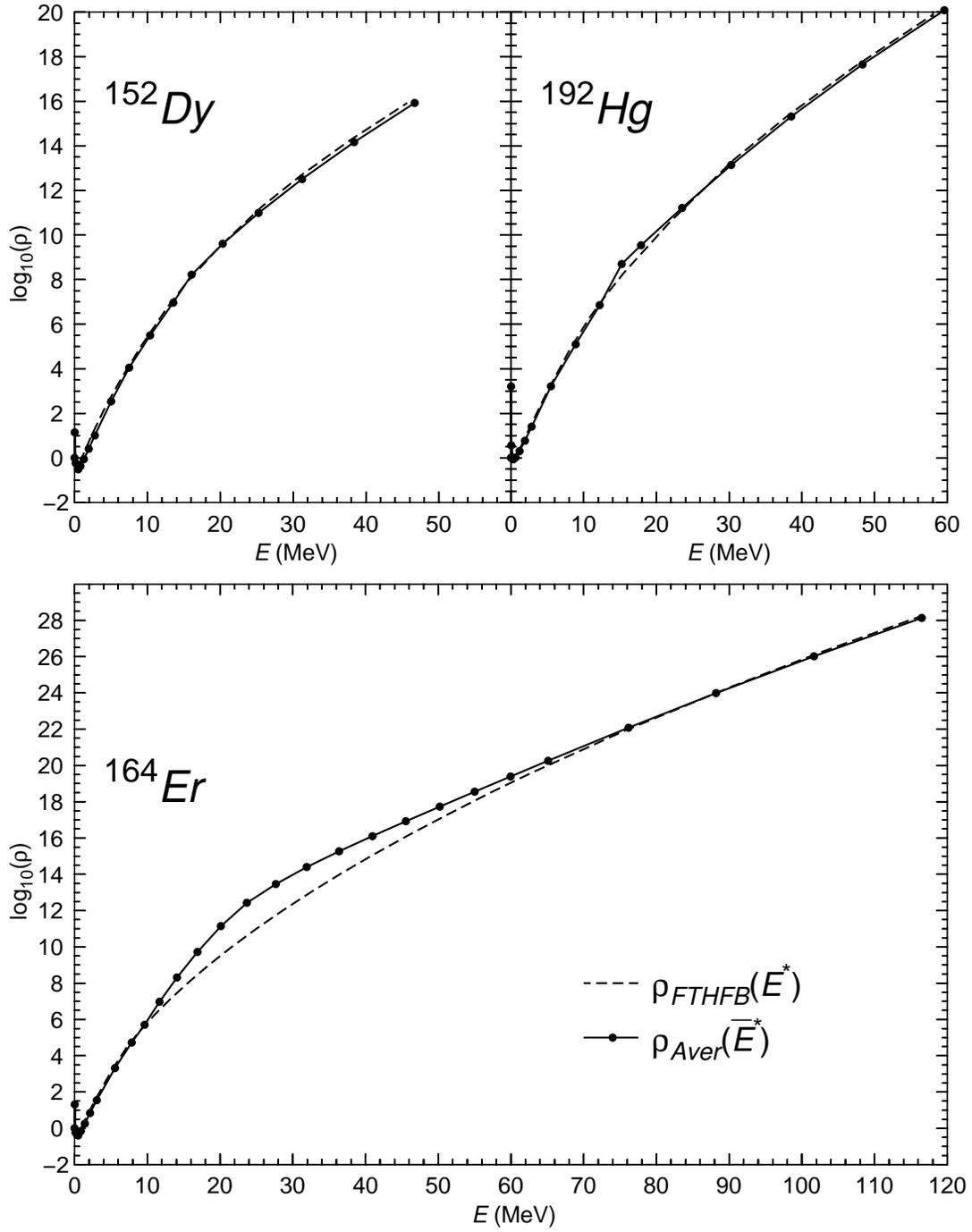}} \par}
\caption{ Total level densities in MeV$^{-1}$ versus the excitation energy for FTHFB (empty symbols)
and averages with shape fluctuations (solid symbols).}
\label{Fig:lden}
\end{figure}

\begin{figure}
{\centering {\includegraphics{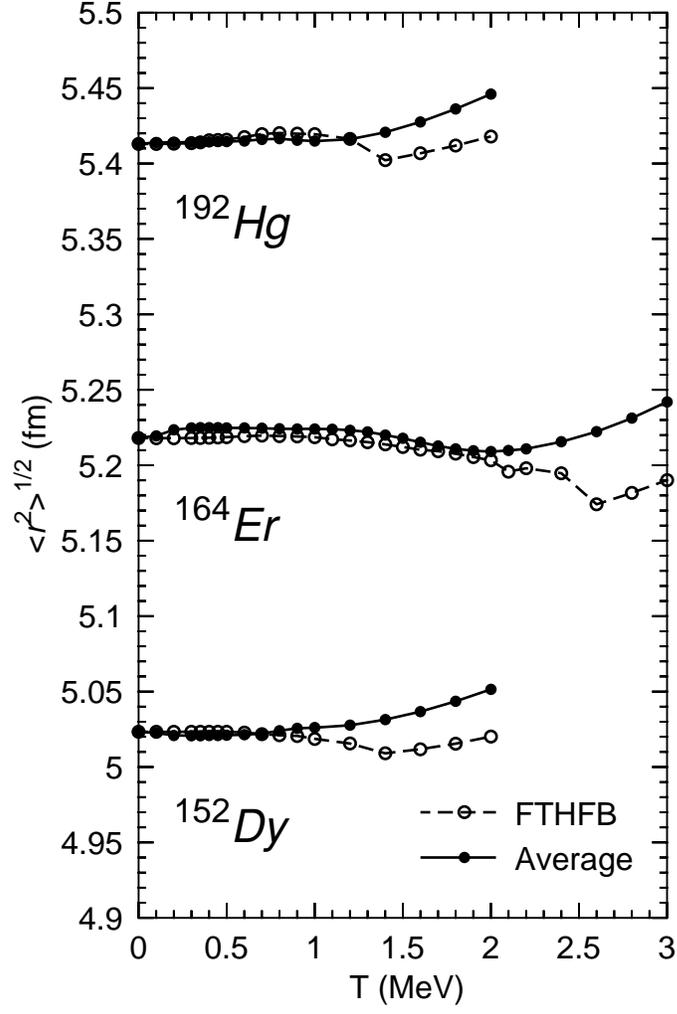}} \par}
\caption{ Root mean squared radii, versus the temperature for FTHFB (empty symbols)
and averages with shape fluctuations (solid symbols). }
\label{Fig:msr}
\end{figure}

\end{widetext}

\end{document}